\documentclass[tradiabstract]{aa}
\usepackage{graphicx}
\usepackage{amsbsy}
\usepackage{amsmath}
\usepackage{natbib}
\usepackage{color}
\usepackage{txfonts}
\usepackage{url}
\usepackage{bm}

\DeclareMathSymbol{\varOmega}{\mathord}{letters}{"0A}
\DeclareMathSymbol{\varSigma}{\mathord}{letters}{"06}
\DeclareMathSymbol{\varPsi}{\mathord}{letters}{"09}
\DeclareMathSymbol{\varPhi}{\mathord}{letters}{"08}
\DeclareMathSymbol{\varGamma}{\mathord}{letters}{"00}

\begin{document}

\title{Anatomy of rocky planets formed by rapid pebble accretion \\ III.\
Partitioning of volatiles between planetary core, mantle,\\and atmosphere}
\titlerunning{Anatomy of rocky planets formed by rapid pebble accretion III}

\author{Anders Johansen\inst{1,2}, Thomas Ronnet\inst{2}, Martin
Schiller\inst{1}, Zhengbin Deng\inst{1} \& Martin Bizzarro\inst{1}}
\authorrunning{Johansen et al.}

\institute{$^1$ Center for Star and Planet Formation, GLOBE Institute,
University of Copenhagen, \O ster Voldgade 5-7, 1350 Copenhagen, Denmark \\ $^2$
Lund Observatory, Department of Astronomy and Theoretical Physics, Lund
University, Box 43, 221 00 Lund, Sweden, \\e-mail:
\url{Anders.Johansen@sund.ku.dk}}

\date{}

\abstract{Volatile molecules containing hydrogen, carbon, and nitrogen are key
components of planetary atmospheres. In the pebble accretion model for rocky
planet formation, these volatile species are accreted during the main planetary
formation phase. For this study, we modelled the partitioning of volatiles
within a growing planet and the outgassing to the surface. The core stores more
than 90\% of the hydrogen and carbon budgets of Earth for realistic values of
the partition coefficients of H and C between metal and silicate melts. The
magma oceans of Earth and Venus are sufficiently deep to undergo oxidation of
ferrous Fe$^{2+}$ to ferric Fe$^{3+}$. This increased oxidation state leads to
the outgassing of primarily CO$_2$ and H$_2$O from the magma ocean of Earth.  In
contrast, the oxidation state of Mars' mantle remains low and the main outgassed
hydrogen carrier is H$_2$. This hydrogen easily escapes the atmosphere due to
the irradiation from the young Sun in XUV wavelengths, dragging with it the
majority of the CO, CO$_2$, H$_2$O, and N$_2$ contents of the atmosphere. A
small amount of surface water is maintained on Mars, in agreement with proposed
ancient ocean shorelines, for moderately low values of the mantle oxidation.
Nitrogen partitions relatively evenly between the core and the atmosphere due to
its extremely low solubility in magma; the burial of large reservoirs of
nitrogen in the core is thus not possible. The overall low N contents of Earth
disagree with the high abundance of N in all chondrite classes and favours a
volatile delivery by pebble snow. Our model of rapid rocky planet formation by
pebble accretion displays broad consistency with the volatile contents of the
Sun's terrestrial planets.  The diversity of the terrestrial planets can
therefore be used as benchmark cases to calibrate models of extrasolar rocky
planets and their atmospheres.}


\maketitle

\section{Introduction}

The delivery of volatiles to the terrestrial planets in the Solar System is
traditionally ascribed to impacts of primitive, water-rich asteroids that
originated beyond the water ice line
\citep{Morbidelli+etal2000,RaymondIzidoro2017}. In this view, the amounts of
volatiles such as carbon, nitrogen, and hydrogen are inherently stochastic and
could be sourced from a wide range of parent bodies with different isotopic
compositions \citep{Marty2012,Alexander+etal2012}. This gives the possibility to
infer the source population of Earth's volatiles by comparing, for example, the
measured D/H ratio of Earth to that of various meteorite groups.

The atmospheres of Venus, Earth, and Mars clearly do have very different
atmospheric volatile contents today. Venus hosts a dense and massive atmosphere
dominated by 96.5\% CO$_2$ with an approximate mass of $M_{\rm CO_2} = 4.6
\times 10^{20}\,{\rm kg}$ (equivalent of $35\%$ of the total water mass in the
Earth's oceans). Its atmosphere was likely outgassed from the mantle in
periodic, global resurfacing events and subsequently trapped because of the lack
of surface water reservoirs for carbonate formation and the reburial of carbon
into the mantle \citep{Taylor+etal2018}. The N$_2$ budget of Venus' atmosphere
is about three times that of Earth's atmosphere, but Earth likely aintains
an additional few parts-per-million (ppm) of N$_2$ buried in the mantle to make
the reservoir more equal to Venus'
\citep{Marty2012,CartignyMarty2013,Mysen2019}.  Earth, in contrast to Venus,
holds much more carbon in the mantle than in the atmosphere
\citep{DasguptaHirschmann2010,Wong+etal2019}, with an estimated range of total
carbon mass between $M_{\rm C}$$\sim$$10^{20}{-}10^{21}$ kg; the lower end
corresponds to the Venusian atmosphere reservoir (using the molecular mass ratio
$m_{\rm CO_2} \approx 3.7 m_{\rm C}$).

Modern Mars has a very tenuous CO$_2$ atmosphere with a pressure of just 0.006
bar. Together with the frozen component at the polar ice caps, the total CO$_2$
reservoir mass of Mars is around $M_{\rm CO_2} \approx 3 \times 10^{16}\,{\rm
kg}$ \citep{Kelly+etal2006}, which is at least three orders of magnitude below
the reservoirs inferred for Earth and Venus. Mars likely experienced extensive
loss of atmospheric volatiles due to XUV irradiation from the young Sun (i.e.\
radiation emitted at X-ray and UV wavelengths) as well as continuous solar wind
stripping \citep{Erkaev+etal2014,Jakosky+etal2015,Wordsworth2016a}. The current
global equivalent layer of water ice on Mars is estimated to be only 20--40 m in
the surface reservoirs \citep{CarrHead2015}. Combined geological evidence, loss
rate calculations, and D/H measurements yield estimates of a primordial water
reservoir in the range between 100 and 1,500 metres \citep{Scheller+etal2021}.
Ancient ocean shore lines \citep{diAchilleHynek2010} indicate a primordial
reservoir of approximately 500 m of water. The latter corresponds to a total
primordial water mass of approximately $M_{\rm H_2O} = 7 \times 10^{19} \,{\rm
kg}$, which is equivalent to 10\% of the Earth's modern oceans.

At face value, the very different volatile inventories of the terrestrial
planets could be interpreted as in agreement with the stochastic volatile
delivery model discussed above. However, both atmospheric escape as well as the
partitioning of volatiles between the atmosphere, the mantle and the core must
have played significant roles in shaping the modern atmospheric reservoirs. For
example, the oxidation state of the bulk mantle has a decisive effect on the
composition of the atmosphere outgassed from the magma oceans of young
terrestrial planets. Earth and Venus are massive enough to have undergone
oxidation processes where ferrous iron Fe$^{2+}$ is oxidized to ferric iron
${\rm Fe}^{3+}$ at the high pressures in their deep magma oceans
\citep{Armstrong+etal2019,Deng+etal2020a}. This leads to the outgassing of an
early atmosphere primarily consisting of CO$_2$ and H$_2$O
\citep{Sossi+etal2020}, while the lower oxidation state of the martian mantle
would have led to the outgassing of huge amounts of H$_2$ that drove an early,
hydrodynamic escape of both H- and C-bearing atmospheric species.

The molecular carriers of H, C, and N are either moderately (H$_2$O) or highly
(CO$_2$ and N$_2$) siderophile, meaning that they partition preferentially
into the metal melt during the magma ocean phase of terrestrial planet
formation \citep{Li+etal2020,Fischer+etal2020,Grewal+etal2019a}. The relatively
similar reservoirs of N (at the $\sim$1 ppm level over the full planetary mass)
in the atmospheres of Earth and Venus is intriguing. This level is an stark
contrast to some of the most commonly assumed source material for Earth, namely
the enstatite chondrites containing 100 ppm of (refractory) N, the carbonaceous
chondrites containing up to 1,000 ppm of (volatile) N and the ordinary
chondrites containing 1--30 ppm of N \citep{Grewal+etal2019a}. Nitrogen does
partition into metal melt \citep{Grewal+etal2021}, but its solubility in magma
is very low to begin with \citep{Sossi+etal2020}, so the core must compete with
atmospheric outgassing \citep{Speelmanns+etal2019}. The low contents of nitrogen
on Earth and Venus could thus be a smoking gun of pebble accretion, since
thermal processing of pebbles in the gas envelope of a growing terrestrial
planet will limit the accretion of the most volatiles species to the earliest
accretion stages \citep{Johansen+etal2021}.

The goal of this paper is to demonstrate that the volatile inventories of Venus,
Earth, and Mars can be explained within the single framework of pebble accretion
and volatile delivery via pebble snow \citep{Ida+etal2019,Johansen+etal2021},
without invoking delivery by stochastic impacts. This will allow us to make a
quantitative and predictive model for terrestrial planet formation and the
compositions of the outgassed atmospheres. Hence, we can use the Solar System to
understand also the atmospheres of terrestrial planets and potentially habitable
planets around other stars.

The origin of life on the planetary surface is a complex and multifaceted
problem. The famous Urey-Miller experiments demonstrated how a reducing
atmosphere dominated by molecules such as H$_2$ and CO exposed to energetic
lightning discharges is prone to catalysis of organic molecules
\citep{MillerUrey1959,Miyakawa+etal2002}. However, as discussed above, the early
atmosphere of Earth may have been already oxidizing. Neutral or oxidizing
atmospheres synthesize far lower amounts of organic molecules in lightning
discharge experiments \citep{SchlesingerMiller1983,Cleaves+etal2008}. In
contrast, the reducing atmosphere on early Mars could have been more conducive
to the origin of life \citep{Deng+etal2020b,Liu+etal2021}. However, the loss of
the martian atmosphere is a challenge to Mars as a cradle of life
\citep{Wordsworth2016a}. After the hydrodynamical escape of its primordially
outgassed atmosphere, competition between volcanic outgassing and hydrogen
escape of these gases would have led to intermittent periods of either reducing
or oxidizing atmospheric conditions \citep{Lanza+etal2016,Wordsworth+etal2021}.
Impacts of metal-rich asteroids could also have created temporary reducing
atmospheres on both Earth and Mars until the released hydrogen escaped
\citep{Zahnle+etal2020,CitronStewart2022}.

On our own Earth, it has been proposed that life arose in warm little surface
ponds that experienced periodic wet-dry cycles leading to the assembly of
complex organic molecules and protocells
\citep{DaSilva+etal2015,Hassenkam+etal2020}. In this view, the feedstock organic
molecules were delivered to Earth by comets and meteorites
\citep{Pearce+etal2017}, with the massive early atmosphere helping to reduce the
impact speed and hence the survival of the organic molecules
\citep{ChybaSagan1992}. Alternatively, life on Earth may have originated in
undersea hydrothermal vent systems near mid-ocean ridges
\citep{ShockSchulte1998,MartinRussell2003,Kelley+etal2005}. Understanding the
composition and evolution of the early atmospheres of terrestrial planets and
their water oceans is under all circumstances a key astrophysical and
geochemical deliverable that provides the initial conditions for the prebiotic
chemistry that led to the origin of life.

\begin{figure*} 
  \begin{center}
    \includegraphics[width=0.8\linewidth]{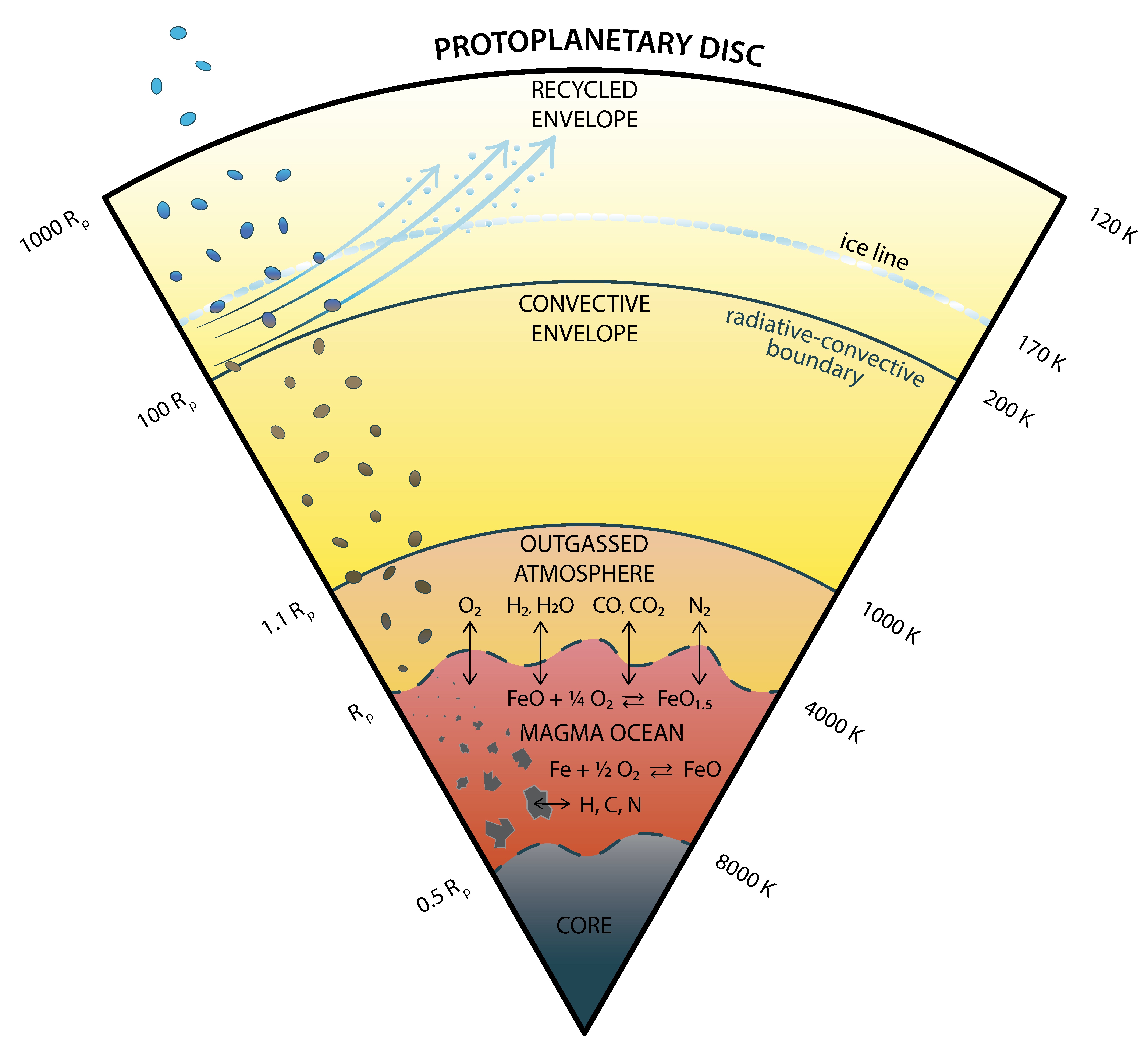}
  \end{center}
  \caption{Overview of the core, mantle, atmosphere and envelope of a rocky
  planet growing by rapid pebble accretion. The protoplanetary disc flows
  penetrate through the recycled part of the envelope, carrying along water
  vapour released at the ice line at $T \approx 170$ K back to the
  protoplanetary disc. The radiative-convective boundary at $T \approx 200$ K
  marks the transition to a bound, convective envelope. The desiccated pebbles
  sediment through the convective gas down to the outgassed atmosphere.  The
  atmosphere is in vapour equilibrium with the silicate magma ocean and pebbles
  therefore survive until they plunge into the magma ocean. Here the metal
  separates from the silicates and metal blobs coalesce and sediment to the
  core. There is an equilibrium between the concentration of volatiles H, C and
  N in the magma ocean and the outgassed atmosphere. The speciation of the
  volatiles in the atmosphere depends on the oxygen fugacity of the magma ocean,
  which is set by the iron-w\"ustite buffer Fe + (1/2) O$_2$
  $\leftrightharpoons$ FeO. The lower regions of the magma ocean undergoes
  additional oxidation by the process FeO + (1/4) O$_2$ $\leftrightharpoons$
  FeO$_{1.5}$. Finally, the volatiles H, C, and N are all siderophile and
  partition primarily into metallic melt over silicate melt. This results in a
  core that holds the main reservoirs of H and C, while N has so low solubility
  in magma that the atmosphere is the dominant reservoir of nitrogen.}
  \label{f:slice_of_planet}
\end{figure*}

The paper is organized as follows. In Section 2 we describe the modules of the
ADAP code dedicated to calculating the escape of the atmospheric constituents
after the dissipation of the protoplanetary disc. In Section 3 we discuss the
fate of H-bearing and C-bearing molecules and how their carriers depend on the
oxidation state of the mantle. In Section 4 we focus instead on nitrogen, since
this element is both siderophile (enters the core) and atmophile (outgasses
easily from the magma ocean). We show that the thermal processing of nitrogen
accreted with pebble snow yields final concentrations that are consistent with
Earth and Venus. In Section 5 we summarize the pebble accretion model for
planetary accretion and differentiation put forward in this paper together with
the companion papers by Johansen et al.\ (2023a,b, hereafter Paper I and Paper
II). Figure \ref{f:slice_of_planet} shows an overview of the core, mantle,
atmosphere and envelope of a rocky planet that grows by our proposed model of
rapid pebble accretion.

\section{ADAP code modules on atmospheric escape}

The interior structure and outgassing modules of the ADAP code are described in
Paper I and Paper II. Here we describe our approach to atmospheric
mass loss after the dissipation of the protoplanetary disc.

\subsection{Atmospheric escape}

We follow \cite{Erkaev+etal2007} and \cite{Salz+etal2016} and define the mass
loss rate of the atmosphere as
\begin{equation}
  \dot{M} = \frac{3 \beta^2 \eta F_{\rm XUV}}{4 K G \rho_{\rm pla}} \, .
  \label{eq:Mdot}
\end{equation}
Here $\beta = R_{\rm XUV}/R_{\rm pla}$ parameterizes the effective capture
radius of XUV photons $R_{\rm XUV}$ relative to the planetary radius $R_{\rm
pla}$, $\eta$ is the efficiency of mass loss relative to the energy-limited
expression (defined as $\eta=1$), $F_{\rm XUV}$ is the energy flux in the XUV
range by the young Sun, $K$ is a parameter of the order of unity that
parameterizes the effect of the stellar gravity on the planetary loss rate and
$\rho_{\rm pla}$ is the internal mass density of the planet.

We assume for simplicity that the mass loss occurs only through direct
acceleration of H atoms and the associated drag by the escaping H on the heavier
species C, O, and N. The intense radiation environment at the base of the
thermosphere is assumed to dissociate all molecules into their constituent atoms
\citep{Erkaev+etal2014}. We define from equation (\ref{eq:Mdot}) an equivalent
number loss rate of hydrogen atoms with of mass $\mu_{\rm H}$ as
\begin{equation}
  \dot{N}_{\rm H}^\star = \frac{3 \beta^2 \eta F_{\rm XUV}}{4 K G \rho_{\rm
  pla} \mu_{\rm H}} X_{\rm H} \, . \label{eq:NdotHstar}
\end{equation}
Here $N_i$ is the total number of atoms of species $i$ (where $i$ denotes the
atomic species H, C, O, and N), $X_i = n_i/n$ is the mixing ratio of species $i$
in the thermosphere, with $n = \sum_i n_i$ denoting the sum of the number
density of all species $n_i$, taking into account that the water mixing ratio
may fall with height due to water cloud condensation (see Paper II). The
multiplication by the hydrogen mixing ratio $X_{\rm H}$ into the number loss
rate in equation (\ref{eq:NdotHstar}) reflect the energy lost to heating of
species other than H.  We thus assume that the absorption of XUV radiation (a)
has a similar cross section for H, C, O, and N atoms and (b) only leads to
escape of H while the heavier species remain bound even after absorption of an
XUV photon. These assumptions allow us to make relatively simple calculations of
the escape of primordial atmosphere dominated by hydrogen.
\begin{figure*}
  \begin{center}
    \includegraphics[width=0.9\linewidth]{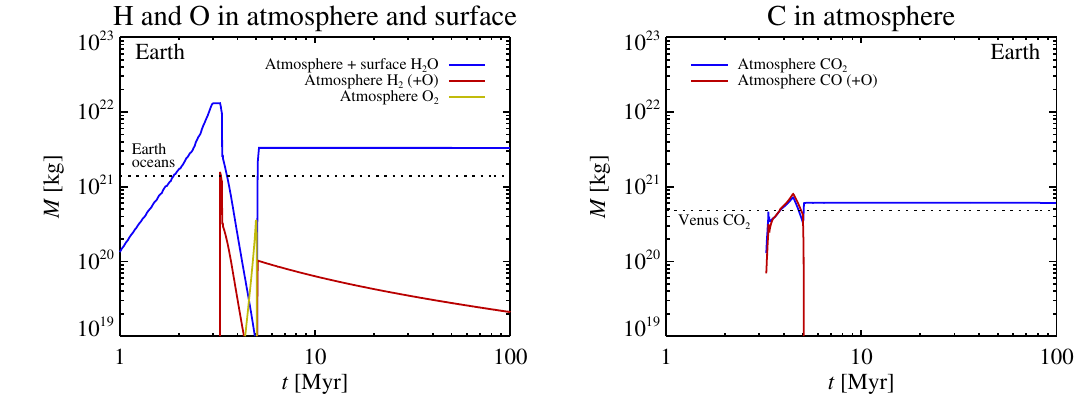}
    \includegraphics[width=0.9\linewidth]{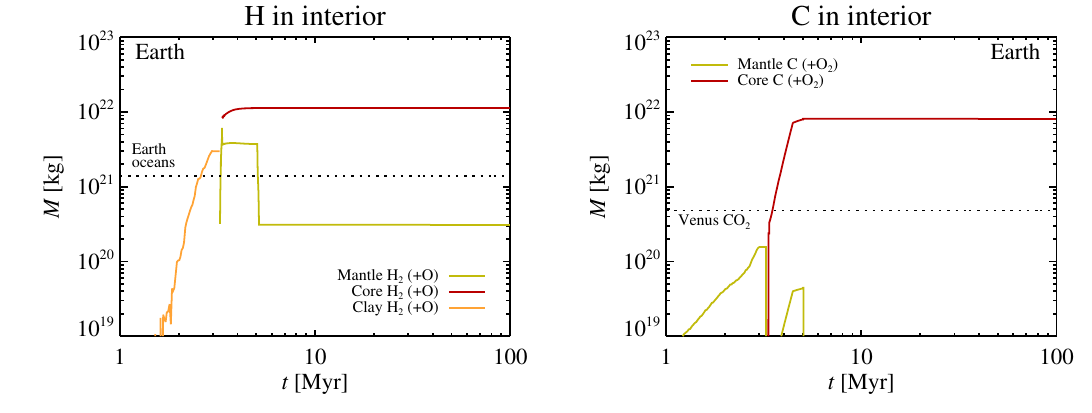}
  \end{center}
  \caption{The evolution of H-bearing (left panels, including O$_2$) and
  C-bearing volatiles (right panels) in the atmosphere and surface (top panels)
  and interior (bottom panels) of our Earth analogue. Hydrogen resides initially
  in a massive surface layer of ice and water and an interior layer of clay.
  The onset of the run-away greenhouse effect after 3.5 Myr heats the surface to
  form a global magma ocean. The water dissolves in the magma and partitions
  within the magma ocean strongly into the metal melt. The core of Earth
  incorporates after the termination of accretion more than 90\% of the total
  hydrogen and carbon budgets. The oxygen fugacity of the magma ocean rises
  with increasing temperature and hence O$_2$ becomes the dominant molecule
  in the atmosphere at the end of the accretion stage; this oxygen is
  nevertheless dissolved back into the magma ocean as it cools. The terrestrial
  magma ocean expels approximately one modern ocean mass of water after
  crystallization of the magma. The oxidized atmosphere of Earth avoids any
  extensive mass loss by XUV irradiation and maintains an atmosphere consisting
  mainly of CO$_2$ with a mass similar to modern Venus' atmosphere.}
  \label{f:atmosphere_evolution_Earth}
\end{figure*}

All the atmospheric components escape together, driven by drag from the hydrogen
atoms but maintaining their initial mixing ratio. We assume that the mixing
ratio in the thermosphere is the same as in the bulk atmosphere, neglecting
any scale-height differences due to the mean molecular weight. This gives a
number loss rate of the mixed species as
\begin{equation}
  \dot{N} = \frac{\dot{N}_{\rm H}^\star}{X_{\rm H} + X_{\rm C}
  (\mu_{\rm C}/\mu_{\rm H}) + X_{\rm O} (\mu_{\rm O}/\mu_{\rm H}) + X_{\rm N}
  (\mu_{\rm N}/\mu_{\rm H})} \, .
  \label{eq:Ndot}
\end{equation}
Here $\mu_i$ is the atomic weight of species $i$. The individual number loss
rates can then be written as
\begin{equation}
  \dot{N}_i = X_i \dot{N} \, ,
  \label{eq:Ndoti}
\end{equation}
which maintains the mixing ratio $\dot{N}_i / \dot{N} = N_i/N_{\rm all}$ for H,
C, O, and N. We furthermore need to ensure that the drag from the escaping H
atoms is sufficient to accelerate the heavier species to the escape speed. This
is seen by calculating the escape parameter
\begin{equation}
  \varPhi_i = \frac{\dot{N}_{\rm H} k_{\rm B} T_{\rm photo}}{3 \pi G M_{\rm pla}
  \mu_i b_{{\rm H},i}} \, .
  \label{eq:Phii}
\end{equation}
Here $T_{\rm photo}$ is the temperature at the photosphere and $b_{{\rm H},i}$
is the binary diffusion coefficient of species $i$ dragged by H
\citep{ZahnleKasting1986,Erkaev+etal2014}. We then set $\dot{N}_i=0$ when
$\varPhi_i < (\mu_i-\mu_{\rm H})/\mu_i$, assuming that the energy transferred
from H to the heavier species in that case goes to heating but does not lead to
escape. This way the H escape flux drives the escape of the heavier species only
when the flux is high, while the heavy species (C, N, O) eventually decouple
from the H drag for lower flux values. The H flux decreases both because the H
mixing ratio of the atmosphere decreases and because the XUV luminosity of the
young Sun falls with time.
\begin{figure*}
  \begin{center}
    \includegraphics[width=0.9\linewidth]{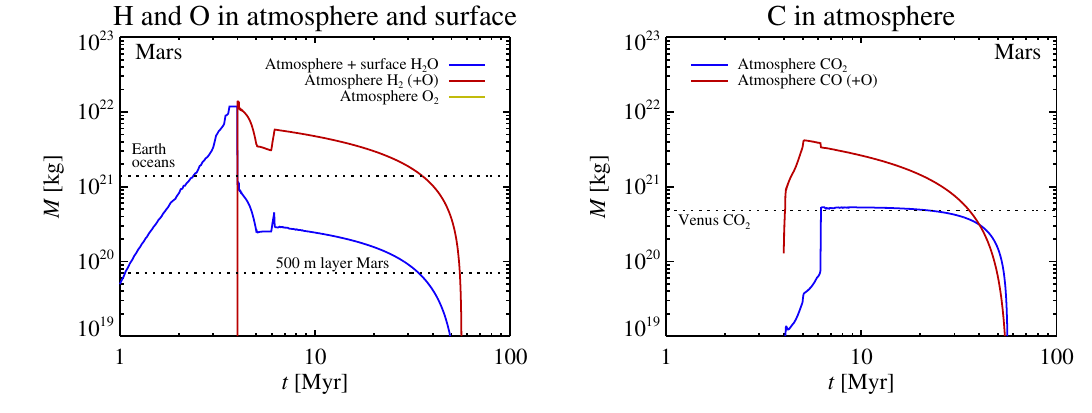}
    \includegraphics[width=0.9\linewidth]{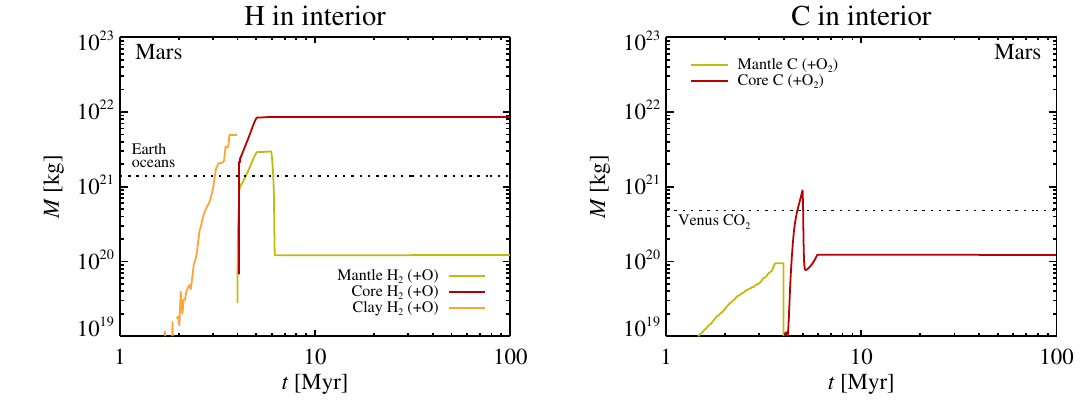}
  \end{center}
  \caption{The evolution of the H-bearing and C-bearing volatiles in our Mars
  analogue.  The smaller martian core can only hold 50\% of the water, resulting
  in the outgassing of a massive H$_2$-dominated atmosphere (we show for H$_2$
  its equivalent water mass). The dominant H$_2$ component of Mars' atmosphere
  undergoes hydrodynamical escape within 70 Myr and drags along with it the
  entire atmospheric CO, CO$_2$ and H$_2$O contents.}
  \label{f:atmosphere_evolution_Mars}
\end{figure*}
\begin{figure}
  \begin{center}
    \includegraphics[width=0.9\linewidth]{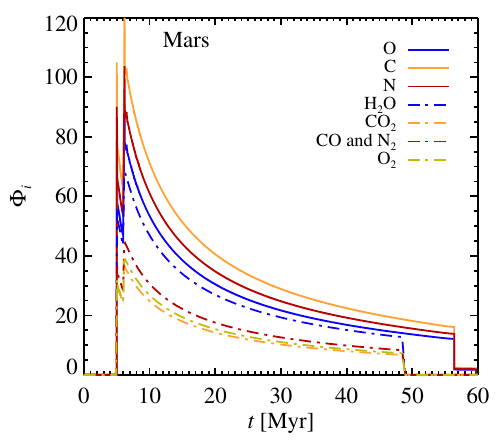}
  \end{center}
  \caption{The escape parameter $\varPhi_i$ as a function of time for our Mars
  analogue. We show results both for our standard assumption that molecules are
  dissociated into O, C, and N in the thermosphere (full lines) as well as for
  an alternative model where the molecules are assumed to remain intact
  (dash-dotted lines). We bundle CO and N$_2$ together in the plot, due to their
  similarity in both binary diffusion coefficient \citep{ZahnleKasting1986} and
  molecular weight; this gives a similar escape parameter. The escape flux of
  hydrogen is in all cases high enough to drag along all heavier species
  ($\varPhi_i \gg 1$) until all the hydrogen has been depleted.}
  \label{f:dragPhi_t}
\end{figure}

To test the sensitivity of our results to the assumption that all molecules
are atomic at the thermosphere, we implemented an additional model where H is
atomic but H$_2$O, CO$_2$, CO, N$_2$, and O$_2$ remain molecular. These heavier
molecules are harder to lift by the escaping hydrogen atoms. We use here
equations (\ref{eq:Ndot}), (\ref{eq:Ndoti}) and (\ref{eq:Phii}) but with
molecular binary diffusion coefficients $b_{{\rm H},i}$ for the relevant
molecules provided in \cite{ZahnleKasting1986}. Due to the lack of data for
molecular oxygen dragged by atomic H in \cite{ZahnleKasting1986}, we set by
analogy $\varPhi_{\rm O_2}=\varPhi_{\rm N_2}$. We ignore the sedimentation of
the heavier species from the thermosphere, since this has only little effect on
the escape rates for our relevant values of $\varPhi_i \gg 1$
\citep{ZahnleKasting1986,Zahnle+etal1990,Erkaev+etal2014}. We refer to
\cite{Hunten+etal1987} and \cite{Wordsworth+etal2018} for detailed analysis of
the hydrodynamical escape problem and the conditions for successful drag of
heavier species.

For the capture radius $\beta$ and the efficiency relative to energy-limited
escape $\eta$ in equation (\ref{eq:NdotHstar}), we follow the parameterizations
fitted to computer simulations by \cite{Salz+etal2016}. The efficiency flattens
out to a relatively constant value of around 30\% for the low-mass planets
considered here, while the capture radius increases with decreasing planetary
mass as the gravitational acceleration decreases, increasing the scale-height of
the atmosphere. For the first 100 Myr of solar evolution, we read off a fit to
the data of \cite{Tu+etal2015} of
\begin{equation}
  L_{\rm XUV} = L_1 (t/{\rm Myr})^{-0.75} \, ,
\end{equation}
with the luminosity at 1 Myr set for simplicity to a nominal value of
$L_1=10^{24}\,{\rm J\,s^{-1}}$. The XUV flux is then calculated as $F_{\rm XUV}
= L_{\rm XUV}/(4 \pi r^2)$ where $r$ is the distance from the star. The decrease
of the luminosity with time becomes steeper after a few hundred 100 Myr of
stellar rotation evolution, but we do not include these later stages of
atmospheric evolution here.

\subsection{Core-powered mass loss}

Core-powered mass loss has been proposed as another important loss mechanism
affecting planetary hydrogen/helium envelopes
\citep{Ginzburg+etal2016,Ginzburg+etal2018}, complementing the loss by XUV
photoevaporation. The mass loss rate of the envelope is here given by the
thermal flux of gas at the sonic radius $R_{\rm s} = G M_{\rm pla}/(2 c_{\rm
s}^2)$ where the envelope ceases to be bound in the absence of an external
pressure \citep{GuptaSchlichting2019}. The sound speed at the photosphere,
$c_{\rm s}$, is set by the effective cooling temperature of the planet.  The
mass loss rate for the core-powered mass loss mechanism is proportional to the
density at the photosphere of the envelope, and this density is in turn
proportional to the total mass of the envelope. Our planets have hydrogen/helium
envelopes of relatively low masses ($M_{\rm env} \sim 10^{19}\,{\rm kg}$ for
Earth $M_{\rm env} \sim 10^{17}\,{\rm kg}$ for Mars, see Paper II). The mass
loss rates for the low-mass envelope cases considered here are thus dominated by
XUV photoevaporation, since that loss rate is independent of the mass of the
envelope and hence very effective on low-mass envelopes. We therefore do not
include core-powered mass loss in the calculations.

\subsection{Stellar irradiation}

The radiative heating of the young planets by the Sun also becomes relevant once
the protoplanetary disc has vanished. We assume that the stellar luminosity was
70\% of the luminosity of the modern Sun and that the planets had the same
albedo as modern-day Venus ($A=0.77$). This high value of the albedo combined
with the faint, young Sun implies that both Earth and Venus condense their water
vapour atmospheres into oceans. Complex atmospheric circulation models and
realistic albedo values are needed to understand whether Venus started off with
a run-away greenhouse atmosphere or whether surface oceans condensed out on
early Venus \citep{Way+etal2016,Turbet+etal2021}. We illustrate the effect of
the surface albedo by performing an additional numerical experiment with a lower
value of the planetary surface albedo, $A=0.3$ representing a relatively
cloud-free planet such as modern Earth.

\section{Partitioning and loss of H-bearing and C-bearing molecules}

We ran simulations for model analogues of Earth, Mars, and Venus using simple
exponential growth rates as in Paper II. Mars and Venus were evolved to their
current masses, while Earth grew to only $0.6\,M_{\rm E}$ within the adopted 5
Myr lifetime of the protoplanetary disc. This is because of the later collision
with the additional planet Theia, assumed here to have a mass of $0.4\,M_{\rm
E}$, to form the Moon. In Paper II we demonstrated that Theia must have had a
mass of at least $0.3\,M_{\rm E}$ to satisfy the low $^{182}$W abundance in the
Earth's mantle after the collision.
\begin{figure*}
  \begin{center}
    \includegraphics[width=0.33\linewidth]{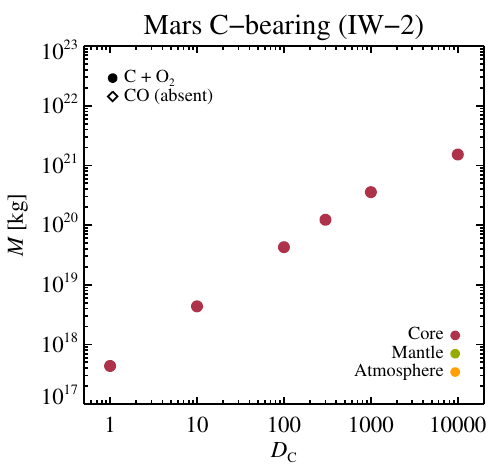}
    \includegraphics[width=0.33\linewidth]{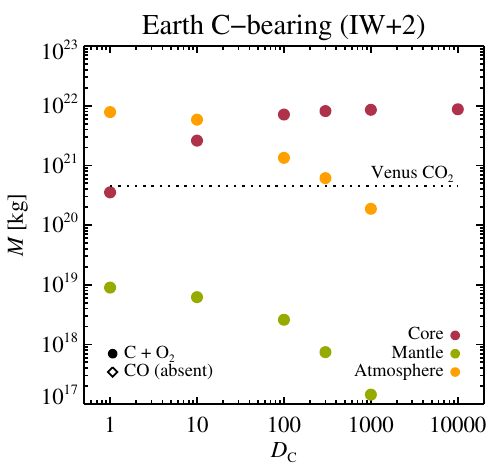}
    \includegraphics[width=0.33\linewidth]{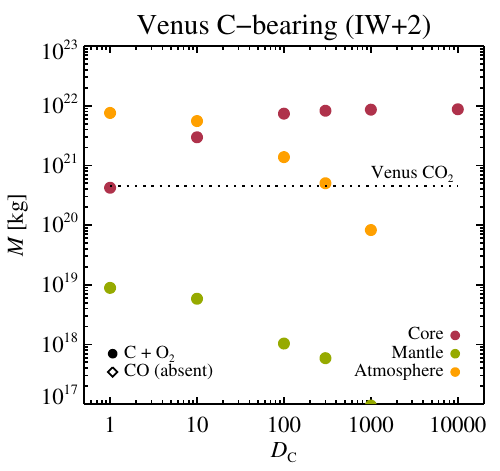}
    \\
    \includegraphics[width=0.33\linewidth]{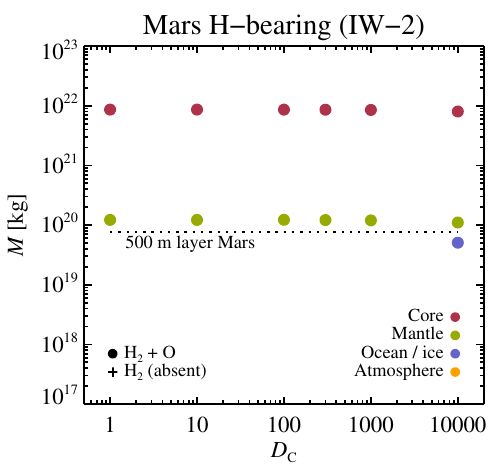}
    \includegraphics[width=0.33\linewidth]{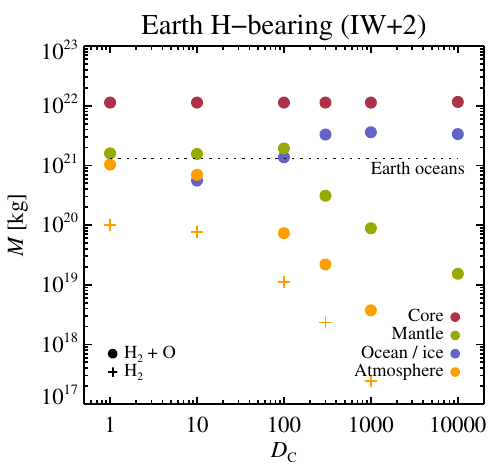}
    \includegraphics[width=0.33\linewidth]{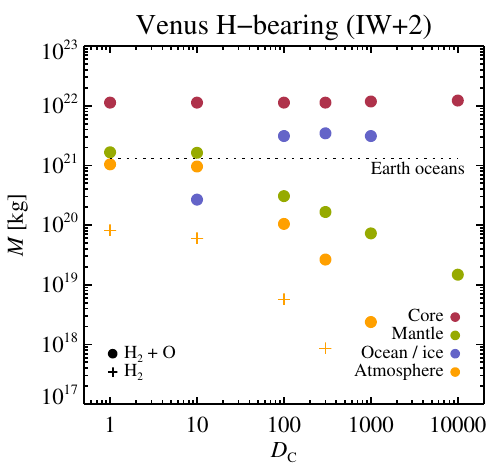}
  \end{center}
  \caption{The distribution of C-bearing (top) and H-bearing (bottom) volatiles
  between core, mantle, ocean, and atmosphere after 100 Myr of atmospheric
  evolution, as a function of the carbon partition coefficient $D_{\rm C}$
  between metal and silicate melt. Mars loses its entire atmospheric reservoirs
  due to atmospheric escape for all values of $D_{\rm C}$. The martian mantle
  nevertheless maintains a significant amount of H$_2$ that can be released as
  water and reduced H$_2$ by later volcanism or impacts and drive short-lived
  warming events. Some ice condenses out at the surface of Mars for the highest
  value of $D_{\rm C}$ where most C is in the core and the atmosphere thus very
  cold. Earth and Venus experience only little atmospheric escape. Their
  atmospheric contents of CO$_2$ match the current atmospheric contents of Venus
  best for a partition coefficient around $D_{\rm C}=300$.  Higher partition
  coefficients bury too much carbon in the core. A partition coefficient around
  our nominal choice of $D_{\rm C}=300$ also gives a good match of the outgassed
  water mass of Earth and Venus to the modern Earth ocean mass.}
  \label{f:composition_Dcar}
\end{figure*}

\subsection{Choice of partition coefficients and oxidation state}

We are particularly interested in understanding the effect of the partition
coefficients of H, C, and N between metal melt and silicate melt,
$D = C_{\rm met}/C_{\rm sil}$ with $C_{\rm met}$ denoting the equilibrium mass
concentration in metal melt and $C_{\rm sil}$ the equilibrium mass concentration
in silicate melt. The measured partition coefficients intrinsically depend on
both temperature and pressure; additionally they come with large experimental
uncertainties and interdependencies between the dissolved species. We therefore
vary the partition coefficients around a set of chosen nominal values. We choose
nominal values $D_{\rm H} = 5$ \citep{Li+etal2020}, $D_{\rm C}=300$
\citep{Fischer+etal2020} and $D_{\rm N}=10$ \citep{Grewal+etal2019a} and examine
how the results depend on varying the partition coefficients around those
values, with $D_{\rm H}$ in the range from 1 to 10, $D_{\rm C}$ in the range
from 1 to 10,000 and $D_{\rm N}$ in the range from 1 to 100. We study a very
large variation in the carbon partition coefficient since the partition
coefficient has been determined experimentally to display a parabolic dependence
on the pressure, with high partition coefficients at both low ($\sim$Mars) and
high ($\sim$Earth) magma ocean pressures \citep{Fischer+etal2020}.

The outgassing of O$_2$ from the magma ocean is key to determining the
speciation between H$_2$/H$_2$O and CO/CO$_2$ in the outgassed atmosphere.  We
choose nominal oxidation states of the mantles of Earth and Venus as IW+2 (where
IW denotes the iron-w\"ustite buffer Fe + (1/2) O$_2$ $\leftrightharpoons$ FeO,
see Paper II), while we assume a much more reduced mantle two log-units below
the buffer (IW-2) for Mars \citep{Armstrong+etal2019,Ortenzi+etal2020}. The loss
of O by atmospheric escape should lead to a reduction of the oxygen fugacity. We
proposed in Paper I that this reduction could be significant for a small body
like Vesta. However, planetary-mass bodies have enormous oxygen reservoirs bound
as FeO in their mantles. We calculate that Mars would have to lose of the order
of 10 Earth oceans and Earth and Venus would have to lose of order the order of
100 Earth oceans to affect the total oxygen budget of mantle plus atmosphere.
We therefore do not change the oxygen fugacity due to atmospheric loss.

\subsection{Evolution of the outgassed atmosphere}

In Figures \ref{f:atmosphere_evolution_Earth} and
\ref{f:atmosphere_evolution_Mars} we show the evolution of H-bearing, C-bearing
species as well as O$_2$ for our Earth and Mars analogues. During the
earliest accretion stages, hydrogen resides mainly as water in a massive surface
ocean and bound as OH in the clay layer below. As the water ocean undergoes a
run-away greenhouse effect after 3.5--4 Myr, the water enters first the
atmosphere as steam and is subsequently dissolved in the newly formed magma
ocean. For Earth, the magma ocean transfers more than 90\% of the water into the
core-forming metal. Approximately one Earth ocean mass of water is outgassed
following the termination of accretion after 5 Myr, leaving only small amounts
of water in the mantle \citep{Elkins-Tanton2008}.

The magma ocean has a low ability to dissolve CO$_2$, so the carbon in our Earth
analogue distributes mainly between the core (holding 90\% of the carbon) and
the atmosphere\footnote{The solubility of carbon actually first increases with
depth in the magma ocean and then decreases at the pressures below 500 km where
diamonds crystallize \citep{Hirschmann2012,Armstrong+etal2019}. The overall
concentration of carbon (in CO$_2$ or diamond) is nevertheless dictated at the
surface interface and maintained throughout the magma ocean by convection.}.
The outgassed CO$_2$ forms a dense greenhouse atmosphere with a temperature of
$T_{\rm surf} = 500$ K. This is cool enough to condense a global water ocean
which would subsequently re-bury CO$_2$ in the mantle after precipitation of
dissolved CO$_2$ to the ocean floor as carbonates
\citep{SleepZahnle2001,Taylor+etal2018}.

Mars initially undergoes a similar evolution to Earth, but its lower mantle
oxidation state results in the outgassing of large amounts of H$_2$. The
hydrodynamical escape of H drives a catastrophic atmospheric mass loss during
the first 70 Myr of stellar activity evolution, leading to the escape of the
entire atmosphere \citep{Erkaev+etal2014}. The mantle of Mars stores $(2-3)
\times 10^{16}\,{\rm kg}$ of CO$_2$, which is approximately equal to the modern
CO$_2$ reservoirs of Mars. The martian mantle also stores H$_2$ equivalent of
approximately 0.1 terrestrial ocean mass if oxidized. However, this hydrogen
storage would be degassed as 10\% H$_2$O and 90\% H$_2$ in volcanism and
impacts, due to the low oxygen fugacity of the martian mantle. In this picture
of early loss of the primordial CO$_2$ atmosphere followed by a gradual resupply
from the mantle, the inferred flow of liquid water on Mars likely took place
during short-lived heating events caused by impacts that released large amounts
of gases such as H$_2$ that acted as temporary thermal blanketing
\citep{Wordsworth+etal2017,Deng+etal2020b}.

\subsection{The efficiency of atmospheric escape}

In Figure \ref{f:dragPhi_t} we show the escape parameter $\varPhi_i$ of
atoms and molecules dragged by atomic hydrogen (defined in equation
\ref{eq:Phii}).  The hydrogen flux is unable to lift the heavier atoms and
molecules when $\varPhi_i < (\mu_i-\mu_{\rm H})/\mu_i \approx 1$, where $\mu_i$
is the mass of atom or molecule $i$. The escape parameter is much larger than
unity for both atoms and molecules in our Mars analogue and therefore the
atmosphere undergoes total escape without leaving heavier species behind. This
happens irrespective of whether or not we assume all molecules to be dissociated
to atoms in the thermosphere.
\begin{figure}
  \begin{center}
    \includegraphics[width=0.9\linewidth]{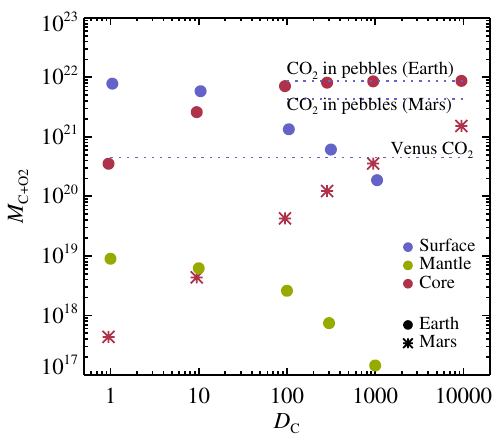}
  \end{center}
  \caption{The mass of CO$_2$ in the surface, mantle and core reservoirs after
  100 Myr of evolution for Earth (filled circles) and Mars (asterisks) as a
  function of the partition coefficient of carbon between metal melt and
  silicate melt. The measurement is given for C + O$_2$ to reflect the CO$_2$
  equivalent mass of the dissolved C in the core and the mantle. We mark the
  accreted CO$_2$ for Earth and Mars as well as the modern CO$_2$ reservoir in
  Venus' atmosphere. Lowering the partition coefficient leads to a decreased
  carbon storage in the core and more outgassing to the surface. The mantle
  holds negligible carbon in all cases.  The escape of the reduced martian
  atmosphere removes any outgassed CO or CO$_2$ reservoirs from Mars. The
  symbols are displaced slightly horizontally to better distinguish them.}
  \label{f:MCO2_DC}
\end{figure}

The partial pressure of oxygen is fixed in our model by the assumption that the
magma ocean works as an oxygen buffer with a fixed fugacity
\citep{Ortenzi+etal2020}. \cite{Hirschmann2020} proposed that the outgassing and
escape of H$_2$ (or correspondingly, the ingassing of O$_2$ from thermal
destruction of H$_2$O in the atmosphere) could increase the oxidation state in
the martian mantle and lead to outgassing of more oxidized species.  This view
nevertheless does not take into account that the oxygen released in the
atmosphere could instead undergo atmospheric escape together with the hydrogen
and hence maintain the mean planetary oxidation state if the escape occurs at a
relative rate of ${\rm O/H}=0.5$.

\subsection{Varying the partition coefficient of carbon}

The adopted value of the partition coefficient of carbon between metal and
silicate melt, $D_{\rm C}$, has key influence on the outgassed atmosphere. In
Figure \ref{f:composition_Dcar} we show the reservoirs of water CO$_2$, H$_2$,
and CO in the core, mantle, ocean, and atmosphere after 100 Myr of atmospheric
evolution. The carbon reservoirs on Earth and Mars are compared directly in
Figure \ref{f:MCO2_DC} as a function of the partition coefficient $D_{\rm C}$.
We show the results for a range of $D_{\rm C}$ values between 1 and 10,000. The
atmosphere of Mars is not much affected by this choice, since the atmosphere is
lost under all circumstances.  The amount of carbon trapped in the core becomes
significant above $D_{\rm C}=300$, but 90\% of the carbon is lost to atmospheric
escape even for large values of $D_{\rm C}$. The cores of Earth and Venus, in
contrast, hold much larger reservoirs of C, unless the $D_{\rm C}$ value is 10
or lower. Very low values of $D_{\rm C}$ result in an extremely massive CO$_2$
atmosphere of more than 10 times the modern Venus atmospheric mass.

The surface water reservoir of Mars (in the form of a global ice layer) is only
significant in the bottom-left panel of Figure \ref{f:composition_Dcar} for the
highest value of $D_{\rm C} = 10000$ where the blue dot approaches 500 metres of
surface ice. This assumption stores the most C in the core and therefore the
atmosphere is colder. Some water is thus maintained during atmospheric escape as
the thinning atmosphere cools to condense out a global ice layer of
approximately 500 m depth. For lower values of $D_{\rm C}$, the ice reservoir on
Mars becomes absent. There is nevertheless significant water stored in mantle
minerals for all values of the carbon partition coefficient (green dots).

Venus and Earth instead hold approximately 2.5 modern ocean masses of water in
their atmospheres and ocean. The balance shifts from ocean to atmosphere below
$D_{\rm C}=100$, as the greenhouse effect from the CO$_2$ is then substantial
enough to keep the temperature above the critical point of water. Hence, Earth
and Venus evolve very similarly. However, this is under the luminosity of the
faint young Sun, set here to 70\% of the modern value. Venus is therefore
expected to undergo a run-away greenhouse effect involving dissociation and loss
of water as the Sun increased its luminosity on the main sequence.
\begin{figure}
  \begin{center}
    \includegraphics[width=0.9\linewidth]{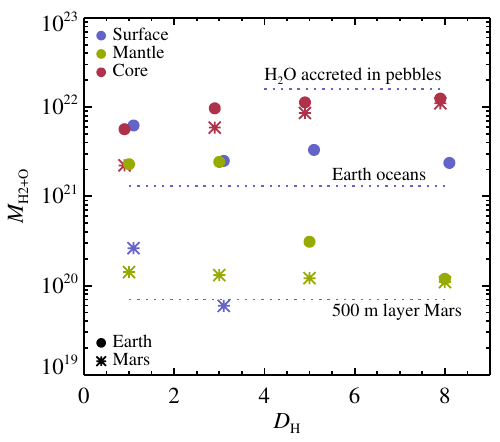}
  \end{center}
  \caption{The mass of water in the surface (condensed and vapour), mantle, and
  core reservoirs after 100 Myr of evolution for Earth (filled circles) and Mars
  (asterisks) as a function of the partition coefficient of water between metal
  melt and silicate melt. The measurement is given for H$_2$+O to reflect the
  H$_2$O equivalent mass of the dissolved H in the core and the mantle.  The
  best match to the surface water reservoir of Earth is found for high values of
  $D_{\rm H}=5$ and $D_{\rm H}=8$. Mars, on the other hand, loses all its
  surface water unless $D_{\rm H}=3$ or less. The core reservoir mass falls with
  decreasing $D_{\rm H}$, particularly for Mars that has a smaller core than
  Earth. This hydrogen was outgassed as H$_2$ from the martian mantle and
  subsequently lost to hydrodynamical escape. The mantles of Earth and Mars
  store significant H$_2$ reservoirs that can later be outgassed as H$_2$ and
  H$_2$O in volcanism and after impacts. The symbols are displaced slightly
  horizontally to better distinguish them.}
  \label{f:Mwat_Dwat}
\end{figure}
\begin{figure}
  \begin{center}
    \includegraphics[width=0.9\linewidth]{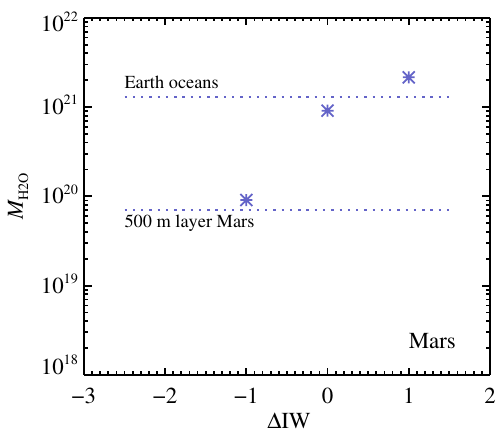}
  \end{center}
  \caption{The mass of water in the surface reservoir of Mars (condensed and
  vapour) as a function of the oxidation state of the mantle relative to the
  iron-w\"ustite buffer (IW), with higher values of $\Delta{\rm IW}$ indicating
  increasing partial pressure of oxygen in the atmosphere. The top line marks
  the modern Earth ocean mass, while the bottom line marks a 500 metre water
  layer on Mars. For the nominal value of $\Delta{\rm IW}-2$, Mars loses its
  entire surface and atmosphere water reservoir, but the water amount increases
  rapidly from IW-1 and upwards.}
  \label{f:Mwat_DIW}
\end{figure}

\subsection{Varying the partition coefficient of water}

The partition coefficient of water between metal melt and silicate melt also
affects strongly the distribution of water between the atmosphere, mantle, and
core. In Figure \ref{f:Mwat_Dwat} we show the mass of water in the three
reservoirs for Earth and Mars, as a function of $D_{\rm H}$ that we vary between
1 and 8, bracketing the range of values reported in \cite{Li+etal2020}.  As
$D_{\rm H}$ is lowered, progressively more water moves from the core to the
atmosphere. The best consistency with the Earth's surface reservoir is found for
high values of $D_{\rm H}=5$ (2.6 oceans) and $D_{\rm H}=8$ (1.8 oceans). Mars
only maintains surface water for $D_{\rm H} \leq 3$, which is much lower than
the nominal value for the partition coefficient.

\subsection{Varying the oxidation state of Mars}

We assumed a nominal oxidation state of the mantle of Mars of ${\rm IW}-2$,
which is valid for a completely open contact between the metal in the core and
the FeO in the mantle \citep{Frost+etal2008} with no further self-oxidation
taking place at high pressures \citep{Armstrong+etal2019}. This value clearly
leads to complete loss of the water from the martian surface reservoir. To
understand how the results depend on the value of $\Delta{\rm IW}$, we ran
simulations for oxidation states between ${\rm IW}-2$ and ${\rm IW}+1$. The
results are shown in Figure \ref{f:Mwat_DIW}. The surface water reservoir
increases steeply when going from very low oxidation states towards ${\rm
IW}+0$. Beyond ${\rm IW}+1$, most of the hydrogen is outgassed from the magma
ocean as H$_2$O and the total amount of water in the surface reservoir
saturates. The best match to Mars' estimated primordial surface water reservoir
comes at an oxygen fugacity slightly above ${\rm IW}-1$. This increase over the
nominal value could indicate that Mars' magma ocean FeO was slightly decoupled
from the core material when the atmosphere was outgassed.
\begin{figure}
  \begin{center}
    \includegraphics[width=0.9\linewidth]{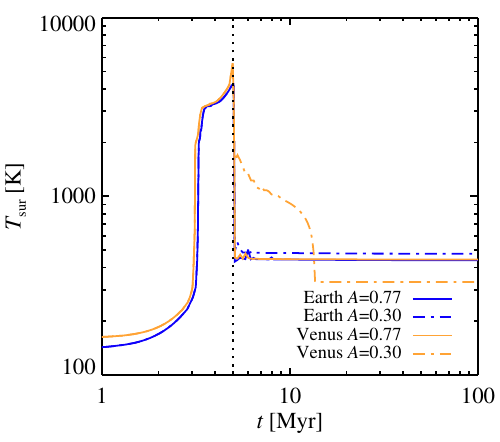}
  \end{center}
  \caption{The surface temperature as a function of time for our Earth and Venus
  analogues evolved for two different values of the planetary albedo. The Earth
  analogue experiences only a slight delay in the cooling when considering the
  lower albedo value, while the cooling of Venus surface is delayed for more
  than 10 Myr. The prolonged state of high surface temperature leads to
  efficient escape of dissociated H$_2$O and CO$_2$; thus the surface
  temperature of Venus eventually lands at a lower value than Earth's.}
  \label{f:Tsur_t_albedo}
\end{figure}

\subsection{Varying the planetary albedo}

Finally, we compare the surface temperature evolution of Earth and Venus for two
values of the planetary albedo: $A=0.77$ (our nominal choice, representing
modern Venus with reflective clouds) and $A=0.3$ (which represents better a
relatively cloud-free planet such as modern Earth). The results are shown in
Figure \ref{f:Tsur_t_albedo}. The high surface temperatures experienced during
the accretion phase are caused by the blanketing effect of CO$_2$ and H$_2$O.
Earth cools down very quickly after the termination of the accretion at
$t=5\,{\rm Myr}$. Venus, in contrast, evolves very differently for the
low-albedo case. Here the cooling of the surface is delayed by more than 10 Myr.
The vicinity of Venus to the Sun in combination with low albedo maintains a
surface temperature of around 1,000 K. Interestingly, the surface temperature
eventually drops as the water vapour in the hot photosphere is easily driven
away by the XUV radiation from the young star. This drags along with it
significant amounts of CO$_2$ as well, which leads to a colder planetary surface
than Earth. We nevertheless consider the high-albedo case more realistic, since
this reflects better the reflection of solar radiation by clouds in
high-temperature atmospheres and is based on modern Venus as a template.
\begin{figure*}
  \begin{center}
    \includegraphics[width=0.9\linewidth]{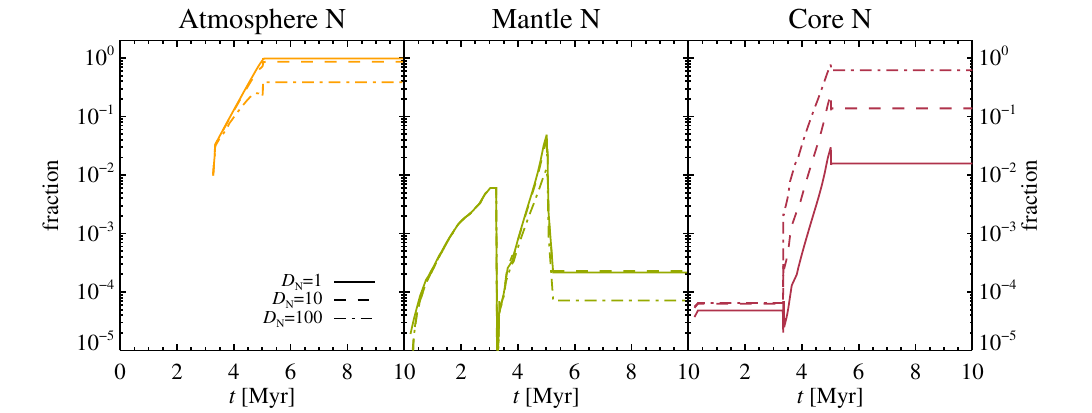}
  \end{center}
  \caption{Evolution of the mass fraction in the N$_2$ reservoirs atmosphere,
  mantle and core for different values of the metal-silicate partition
  coefficient $D_{\rm N}$ in our Earth analogue. We set here the total nitrogen
  mass to an arbitrary value of $3.8 \times 10^{18}\,{\rm kg}$, similar to the
  modern terrestrial atmosphere reservoir. Since N$_2$ in our model does not
  affect the thermodynamics or mass of the planet, this choice is arbitrary. We
  show the results for three values of the metal-silicate partition coefficient,
  $D_{\rm N}=1,10,100$. However, the partitioning between core and mantle has to
  compete also with the outgassing of N$_2$ to the atmosphere. This means that
  even $D_{\rm N}=100$ leads only to approximately equal masses of nitrogen in
  the atmosphere and in the core. Hence it is hard to store significant amounts
  of N in the core of Earth.}
  \label{f:MN2_t}
\end{figure*}

\section{Nitrogen partitioning}

The distribution of nitrogen between the atmosphere and the core is of
particular interest for understanding the source material of Earth. The
atmosphere of Earth contains approximately 0.5 ppm of nitrogen, normalized by
the full mass of the planet. This is in stark contrast to several hundred ppm
found in enstatite chondrites and even higher values in the carbonaceous
chondrites \citep{Grewal+etal2019a}. \cite{Grewal+etal2021} proposed that large
amounts of N are sequestrated in the core of Earth. Transport of nitrogen from
the mantle to the core is expected due to the strong partitioning of N into the
metal melt over the silicate melt. The partition coefficient could be as high as
$D_{\rm N}=10$ or even $D_{\rm N}=100$ for relatively oxidized magma
compositions \citep{Grewal+etal2019a}. However, the transport of N from mantle
to core has to compete with strong tendency for outgassing of nitrogen into the
atmosphere \citep{Speelmanns+etal2019}.
\begin{figure}
  \begin{center}
    \includegraphics[width=0.9\linewidth]{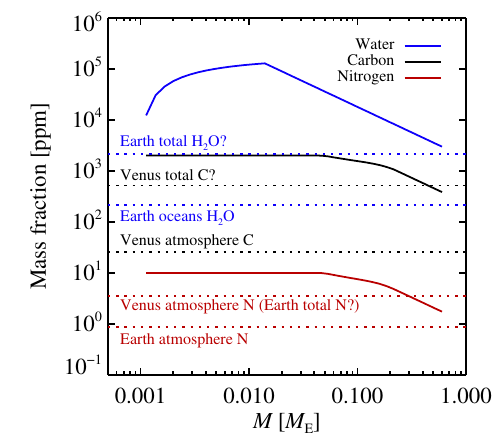}
  \end{center}
  \caption{The mass fraction of water, carbon and nitrogen for the Earth
  analogue model of \cite{Johansen+etal2021} as a function of the accreted mass.
  The dotted lines shows constraints from the volatile reservoirs of Earth and
  Venus, together with estimates of the total water and carbon including the
  core. The volatiles are accreted with pebbles mainly in the early stages of
  planet formation where the envelope temperature is low. We find overall good
  matches to the water and carbon contents of Earth and Venus, as in
  \cite{Johansen+etal2021}. The pebble accretion model also gives a good match
  to the nitrogen reservoirs in the atmospheres of Earth and Venus. In contrast,
  the enstatite chondrites, often considered a possible source material of Earth
  \citep{Dauphas2017}, have N contents above 100 ppm and hence provide a poor
  match to the nitrogen budget of Earth \citep{Grewal+etal2019a}.}
  \label{f:carbon_water_nitrogen}
\end{figure}

\subsection{Modelling nitrogen partitioning}

In addressing the nitrogen problem, we first calculate the relative distribution
of nitrogen between atmosphere, mantle, and core. We assume here that nitrogen
constitutes a constant fraction of the accreted material at an (arbitrary) total
amount corresponding to the atmospheric inventory of the modern Earth. In Figure
\ref{f:MN2_t} we show the fraction of N$_2$ in the different mass reservoirs for
our Earth analogue as a function of time for partition coefficients $D_{\rm
N}=1,10,100$. For low values of $D_{\rm N}$, the majority of the nitrogen ends
in the atmosphere, with only a small amount distributed to the core. Even for
$D_{\rm N}=100$ the core and atmosphere inventories only become about equal;
this is due to the low dissolution of N in the magma melt. Hence even large
values of $D_{\rm N}$ do not lead to a substantially higher core reservoir than
mantle reservoir. This implies then that no chondrites class, except perhaps the
ordinary chondrites, provides the right amount of nitrogen to be reservoirs for
Earth formation. This leaves differentiated bodies that experienced extensive N
outgassing and loss as a possible source planetesimals for Earth formation
\citep{Grewal+etal2019b}. Samples of planetesimals that formed early enough to
melt (e.g.\ Vesta meteorites, ureilites, angrites) nevertheless have isotopic
compositions of lithophile elements such as Cr and Ca that disagree with Earth
\citep{Schiller+etal2018,Zhu+etal2021}. In contrast, if Earth formed mainly by
accreting pebbles, then the nitrogen depletion can be well understood from the
same thermal processing that filtered away water and carbon.

In \cite{Johansen+etal2021} we calculated the final mass fraction of water and
carbon for Earth, using source material where the early accretion phase consists
of 35\% mass of water and 3,000 ppm mass of carbon, while the late-accreted
outer Solar System material carries 5\% carbon \citep{Alexander+etal1998}. This
material is then processed in the hydrostatic hydrogen envelope of the planet,
with the accreted water fraction dropping to zero when the envelope acquires a
temperature above 160 K and carbon sublimated and pyrolized in successive steps
from 325 K to 1,100 K \citep{GailTrieloff2017}. Here we extend this analysis to
nitrogen. We assume that the N contents of the NC reservoir (characteristic of
inner Solar System material) is 10 ppm, while the N contents of the CC reservoir
(characteristic of outer Solar System material) is 500 ppm, the majority of this
nitrogen residing in the organics \citep{Alexander+etal1998}. Nitrogen is then
released as volatile species between 325 K and 425 K, following the sublimation
and pyrolysis of the carbon in the organics
\citep{Nakano+etal2003,GailTrieloff2017}.

\subsection{Nitrogen in core, mantle, and atmosphere}

In Figure \ref{f:carbon_water_nitrogen} we show the mass fraction (in ppm) of
water, carbon, and nitrogen for our Earth analogue as a function of its accreted
mass. We overplot the water ocean reservoir on Earth, the carbon atmospheric
reservoir of Venus and the nitrogen atmospheric reservoirs of Venus and Earth.
\cite{Mysen2019} estimated the mantle reservoir of N on Earth to be twice the
atmospheric value. This way the total N contents of Earth correspond well to the
total mass of atmospheric N on Venus. The release of the bulk N contents of
Venus to its atmosphere could be due to oxidation of the mantle by oxygen left
over from the loss of water in the upper atmosphere \citep{Wordsworth2016b}.
Since significant amounts of water and carbon may be stored in the cores of
Venus and Earth, we also plot estimated values of these total reservoirs.

Figure \ref{f:carbon_water_nitrogen} confirms that the water and carbon
reservoirs of Earth fit well with the pebble accretion model. The nitrogen gives
an equally good fit. The low nitrogen contents of the NC reservoir (at 10 ppm)
combined with thermal processing in the envelope at elevated planetary masses
gives a final mass fraction of nitrogen of around 2 ppm. This is in stark
contrast to the nitrogen contents of the enstatite chondrites, often considered
the source material of Earth \citep{Dauphas2017}. At several 100 ppm of
nitrogen, the enstatite chondrites could at most have provided a percent of the
final masses of Earth and Venus.  The nitrogen isotopes of iron meteorites are
divided in a neutron-poor NC component and a neutron-rich CC component
\citep{Grewal+etal2021}. The isotopic composition of the Earth's nitrogen could
in the pebble accretion picture reflect the accretion of neutron-poor nitrogen
carried by pebbles of inner Solar System composition combined with the late
accretion of a few mass percent of nitrogen-rich planetesimals from the outer
Solar System \citep{Marty2012,Bekaert+etal2020}. Scaling the planetesimal source
material instead to ordinary chondrites, using the abundances from
\cite{Bekaert+etal2020}, yields instead 5--10\% contribution from accretion of
chondrites to Earth. Such planetesimal contribution levels (a few percent CC or
up to 10\% OC) would also deliver reasonable fits to the noble gas abundance of
our planet \citep{Marty2012}.

\section{Summary of pebble accretion model for terrestrial planet formation}

In our three papers on the anatomy of rocky planets forming by pebble accretion,
we have explored a novel model for the formation of terrestrial planets where
the solid mass of the planets as well as the volatiles are delivered mainly via
pebbles (see our overview of the model in Figure \ref{f:slice_of_planet}). This
idea is attractive to explore for three major reasons: (i) observations of
protoplanetary discs reveal large populations of mm-sized pebbles -- planetary
building blocks in the pebble accretion model -- around young stars
\citep{Zhu+etal2019}, (ii) measurements of the isotopic composition of Earth and
the known meteorite classes show that our planet incorporated a large fraction
($\sim$40\%) mass with the composition of material from the outer Solar System
that was likely delivered to our planet via drifting pebbles
\citep{Schiller+etal2018}, and (iii) while the pebble accretion was developed in
order to explain the rapid formation of the cores of gas giants in the outer
Solar System \citep{OrmelKlahr2010,LambrechtsJohansen2012}, the large flux of
pebbles drifting through the terrestrial planet region must by extension have
played a role in the assembly of terrestrial planets as well
\citep{Johansen+etal2021}.

In Paper I we identified the core mass fraction and FeO mantle fraction of
terrestrial planets and planetesimals as a direct consequence of the formation
of the terrestrial planets by pebble accretion exterior of the water ice line.
Water is accreted as ice as long as the envelope is not yet hot enough to
sublimate the ice into water vapour that is recycled with hydrodynamical flows
back to the protoplanetary disc. Upon melting by the accretion heat, the water
oxidizes metallic iron to form magnetite. This oxidized iron in turn becomes
unavailable to contribute to the core. We could therefore show that the
increasing core mass fraction and decreasing FeO mantle fraction of the Vesta,
Mars, Earth triplet is consistent with iron oxidation by early exposure to
liquid water. This in turn implies that the core mass fraction of terrestrial
planets is a monotonously decreasing function of the planetary mass, with the
extrapolated expectation that massive super-Earths should have a very high core
mass fraction if they formed by pebble accretion near the water ice line.

The differentiation of such rapidly formed planets occurs via the continuous
release of accretion energy rather than via giant impacts that melt the mantle,
as we demonstrated in Paper II. The retention of the pebble accretion heat
requires a thermally blanketing atmosphere and here the accreted volatiles play
an important role in outgassing an early atmosphere of H$_2$O and CO$_2$ that
traps the accretion heat. The core of Earth thus forms within the lifetime of
the protoplanetary disc and this would seem at first glance to be in conflict
with the Hf-W dating of core formation, which yields ages in excess of 35 Myr
\citep{KleineWalker2017}. However, we demonstrated in Paper II that a
moon-forming giant impact occurring after 40--60 Myr resets the Hf-W clock and
makes the Earth appear younger than its actual bulk core formation time of only
5 Myr after the formation of the Sun.

Thermal processing of pebbles in the envelope of the growing planets is a key
difference between the pebble accretion model and the traditional view that
volatiles are delivered via planetesimal impacts. Particularly, the terrestrial
planets can grow exterior of the water ice line without accreting excessive
amounts of volatiles \citep{Johansen+etal2021}. Nitrogen appears to be a
decisive volatile in this connection. In solid substances, this atom resides
mainly in organic molecules in the inner regions of the protoplanetary disc and
is hence released successively as the organics sublimate and pyrolyse at
increasing temperatures.  This way we demonstrated in Paper III that the 10 ppm
concentration of nitrogen typical of ordinary chondrite meteorites that formed
in the asteroid belt region is reduced to a few ppm (parts-per-million) by
thermal processing, in agreement with the nitrogen reservoirs residing in the
atmospheres of Earth and Venus. This contrasts strongly to the several hundred
ppm of refractory nitrogen found in enstatite chondrites
\citep{Grewal+etal2019a}, often considered a potential source material for Earth
\citep{Dauphas2017}.

We demonstrated here in Paper III that the surface water reservoir of Earth as
well as the CO$_2$ contents of Venus' atmosphere are reproduced well in the
pebble accretion model using nominal values for the partition coefficients of
these volatiles between metal melt and silicate melt. Preventing complete loss
of the martian surface water requires a slight increase in the oxygen fugacity
of the mantle over the nominal value of ${\rm IW}-2$, indicating perhaps a
gradual decoupling of the magma ocean from the metal in the core. The mantles of
our model Venus, Earth and Mars also store significant hydrogen that can later
be outgassed as H$_2$ and H$_2$O, depending on the evolution of the mantle
oxygen fugacity with time. The cores of the terrestrial planets hold under all
circumstances large reservoirs of H and C that were trapped in the descending
metal droplets during the core-formation stage.

In summary, we believe that we have demonstrated that our pebble accretion model
for terrestrial planet formation displays several key consistencies with the
observed properties of the terrestrial planets in the Solar System. This in turn
implies that these well-studied terrestrial planets can be used as benchmark
cases for the testing of a self-consistent planet formation model that can
subsequently be applied to predict the properties of the interiors and
atmospheres of extrasolar rocky planets as well as the early conditions on
the surface when prebiotic chemistry strives towards ever increasing complexity.

\begin{acknowledgements}

We thank an anonymous referee for carefully reading the three papers in this
series and for giving us many comments and questions that helped improve the
original manuscripts. We also thank the second referee of this paper for many
constructive comments. We are additionally grateful to Ane Hinnum for designing
the overview figure of rocky planet formation by pebble accretion.  A.J.\
acknowledges funding from the European Research Foundation (ERC Consolidator
Grant 724687-PLANETESYS), the Knut and Alice Wallenberg Foundation (Wallenberg
Scholar Grant 2019.0442), the Swedish Research Council (Project Grant
2018-04867), the Danish National Research Foundation (DNRF Chair Grant DNRF159)
and the G\"oran Gustafsson Foundation.  M.B.\ acknowledges funding from the
Carlsberg Foundation (CF18\_1105) and the European Research Council (ERC
Advanced Grant 833275-DEEPTIME). M.S.\ acknowledges funding from Villum Fonden
(grant number \#00025333) and the Carlsberg Foundation (grant number CF20-0209).
The computations were enabled by resources provided by the Swedish National
Infrastructure for Computing (SNIC), partially funded by the Swedish Research
Council through grant agreement no.\ 2020/5-387.

\end{acknowledgements}

\end{document}